**The Community Index: A More Comprehensive Approach to Assessing Scholarly Impact**

Arav Kumar[1],* Cameron Sabet[2],* Alessandro Hammond[3],* Amelia Fiske[4], Bhav Jain[5], Deirdre Goode[6], Dharaa Suresha[7], Leo Anthony Celi[8,9,10]#, Lisa Soleymani Lehmann[11], Ned Mccague[10], Rawan Abulibdeh[12], Sameer Pradhan[13]
* = co-first authors
# = corresponding author
[1] Monroe Township School District, Monroe, NJ, USA
[2] Georgetown University Medical Center, Washington, DC, USA
[3] Harvard University, Cambridge, MA, USA
[4] Institute for the History and Ethics of Medicine Technical University of Munich (TUM) Munich, Germany
[5] Stanford School of Medicine, Palo Alto, CA, USA
[6] Department of Medicine, Brigham and Women's Hospital and Harvard Medical School, Boston, MA USA
[7] Dwight Global Online School, New York, NY, USA
[8] Laboratory for Computational Physiology, Massachusetts Institute of Technology, Cambridge, MA, USA
[9] Division of Pulmonary, Critical Care and Sleep Medicine, Beth Israel Deaconess Medical Center, Boston, MA, USA
[10] Department of Biostatistics, Harvard T.H. Chan School of Public Health, Boston, MA, USA
[11] Department of Health Policy and Management, Harvard T.H. Chan School of Public Health
[12] Department of Electrical and Computer Engineering, University of Toronto, Ontario, Canada
[13] Cemantix.org, Cambridge, MA USA
**Funding:** LAC is funded by the National Institute of Health through DS-I Africa U54 TW012043-01 and Bridge2AI OT2OD032701, the National Science Foundation through ITEST #2148451, and a grant of the Korea Health Technology R&D Project through the Korea Health Industry Development Institute (KHIDI), funded by the Ministry of Health & Welfare, Republic of Korea (grant number: RS-2024-00403047).
**Competing interests:** Authors declare that they have no competing interests.
**Submission Category:** Original Research
**Corresponding Author:**
Leo Anthony Celi, MD
Laboratory for Computational Physiology, Massachusetts Institute of Technology, Cambridge, MA 02139. Division of Pulmonary, Critical Care and Sleep Medicine, Beth Israel Deaconess Medical Center, Boston, MA 02215. Department of Biostatistics, Harvard T.H. Chan School of Public Health, Boston, MA 02115
Email: lceli@mit.edu
ORCID: 0000-0001-6712-6626
Phone number: +1 (617) 253-7818

*Thesis:* This paper aims to create an index to measure the plurality of the community membership of researchers. It is meant to be an additional metric to work alongside the other measurements, like the h-index, with the aim of placing value not just on output but also on a range of collaborative academic relationships. Of note, the relationship between collaboration diversity and research impact depends on how diversity is defined and measured. Moreover, diversity across academic disciplines could enhance impact by hopefully including a broader range of epistemic perspectives in research design and analysis, as well as enabling research to be shared and cited in a wider range of scholarly contexts.

**Abstract:**

The h-index is a widely recognized metric for assessing the research impact of scholars, defined as the maximum value h such that the scholar has published h papers each cited at least h times. While it has proven useful measuring individual scholarly productivity and citation impact, the h-index has limitations, such as an inability to account for interdisciplinary collaboration or demographic differences in citation patterns. Moreover, it is sometimes mistakenly treated as a measure of research quality, even though it only reflects how often a scholar's work has been cited. While metric-based evaluations of research have grown in importance in some areas of academia, such as medicine, these evaluations fail to consider other important aspects of intellectual work, such as representational and epistemic diversity in research. In this article, we propose a new metric called the c-index, or the community index, which combines multiple dimensions of scholarly impact. This is important because a plurality of perspectives and lived experiences within author teams can promote epistemological reflection and humility as part of the creation and validation of scientific knowledge. The c-index is a means of accounting for the often global, and increasingly interdisciplinary nature of contemporary research, in particular, the data that is collected, curated and analyzed in the process of scientific inquiry. While the c-index provides a means of quantifying diversity within research teams, diversity is integral to the advancement of scientific excellence and should be actively fostered through formal recognition and valuation. We herein describe the mathematical foundation of the c-index and demonstrate its potential to provide a more comprehensive representation and more multidimensional assessment of scientific contributions of research impact as compared to the h-index.

## Introduction:

The h-index, introduced by Jorge E. Hirsch in 2005, has become a standard metric for assessing the scholarly impact of researchers[1,2]. It is defined as the maximum value of h such that a researcher has published h papers, each with at least h citations. The two parameters needed to compute the h-index are most likely readily available. The popularity of h-index, possibly owing the two-dimensional parameter space, and its computational ease, by its very definition, fails to capture the plurality of dimensions that comprise the rich landscape of research collaborations and scholarship. Specifically, it does not account for the differences in the nature of research across disciplines, nor does it account for the landscape of increasingly interdisciplinary collaborations, the quality of research produced, cross-geographical research efforts, gender or racial representativeness among authors, or the relative scarcity of contributions from underrepresented specialties. These omissions fail to capture important dimensions of modern research, in particular those within a context where diverse perspectives, backgrounds, and positionalities are needed and valued. Additionally, if utilized as a primary metric in decisions regarding promotion, tenure, and grant funding, the h-index may exacerbate gender and sociodemographic bias, limiting the inclusion and advancement of early- and mid-career marginalized scholars.[3,4] More broadly, the correlation between the h-index and scientific awards has been declining in part due to trends such as hyper-authorship and self-citing.[4] While other indices, such as the g-index, academic trace, integrated impact indicator (I3), R-index, AR-index, International Collaboration Index (ICI) and p-index attempt to address some of these limitations by incorporating factors like citation distribution, collaboration patterns, and publication age, none comprehensively evaluate the representativeness of author research teams across different axes of positionality.[5–9] More specifically they do not include metrics that include demographic, geographic, and epistemological dimensions.[9] The c-index, introduced here, addresses these critical gaps by offering a multidimensional framework for evaluating scholarly impact and placing value on greater diversity of collaboration.

In contrast to existing metrics, the c-index is not meant to replace the h-index as the primary evaluation tool for scholarly impact but rather to complement it. By integrating these additional dimensions, the c-index incentivizes the inclusion of voices that have been historically marginalized in academic publishing, such as women, scholars of color, and those working in underrepresented disciplines or institutions. This plurality of expertise, perspectives and lived experiences fosters more creative research ideas and robust methodologies by enabling a more comprehensive understanding of complex problems and injecting healthy skepticism that is created when different backgrounds and disciplines are interfaced. Greater diversity and greater variation of authors through lived experiences has the ability to catalyze new ideas, discovery, and innovation, and moreover to bring in the perspectives of individuals and communities that have historically been excluded from the ivory tower.

The methodological framework proposed in this paper serves as a starting point by which future iterations of the c-index can be developed in collaboration with marginalized scholars defined as researchers who have been historically underrepresented their race, ethnicity, gender, socioeconomic background, or disability status. By recognizing plurality as an integral part of scholarly impact, the c-index encourages collaborations across disciplines, geographies, and identities. As such, the c-index represents a significant advancement in measuring and promoting inclusivity in research, complementing traditional metrics and fostering a more equitable academic landscape.

**Methods:**

In this section, we formally define the c-index, a metric for evaluating the plurality of the community of a researcher across various dimensions, including gender, country, and academic discipline. To account for the potential inflation of this metric through repeated co-authorships, particularly when such collaborations are concentrated within a narrow domain of inquiry, we introduce a novelty multiplier. This mechanism increases the weight of collaborations with new co-authors by awarding a higher contribution to first-time edges in the author's ego network The c-index is calculated using a graph-theoretical framework, where nodes represent authors and edges represent co-authorship relationships.

Representation of Collaborations

Let $G = (V, E)$ be an undirected graph where $V$ is the set of authors, and $E$ is the set of edges representing co-authorship relationships between authors. For each author $v_i \in V$ we define their neighborhood $N(v_i) \subseteq V$ as the set of co-authors with whom author $v_i$ has collaborated. The size of the neighborhood $|N(v_i)|$ represents the number of distinct co-authors of $v_i$.

Definition of the c-index

The c-index $D(v_i)$ of an author $v_i$ is designed to reflect the plurality of scholarly community engagement across three core dimensions: gender, geography, and disciplinary affiliation. These three dimensions were selected based on longstanding evidence that they represent core axes of diversity in science, shaping epistemic communities, access to resources, and the production of knowledge. Gender and geographic disparities have been widely documented as structural barriers to participation and recognition in scienc[11,45-47], while disciplinary silos continue to constrain the circulation of ideas and interdisciplinary collaboration[48,49]. Together, these dimensions provide a meaningful and generalizable foundation for evaluating the plurality of scholarly communities. While other dimensions of diversity are undoubtedly important (e.g., race, ethnicity, institutional affiliation, and language) the selection of gender, geography, and discipline reflects a pragmatic focus on attributes that are (a) widely documented in the literature as key sources of inequality in science and (b) consistently and systematically available in large-scale bibliometric datasets.

For each author, the index aggregates paper-specific diversity scores and applies an additional adjustment that captures the dynamism of their collaborative network. Specifically, we introduce a *novelty multiplier*, which rewards authors for forming new co-authorship ties over time. For each publication, we assess whether co-authors represent newly added edges in the author's collaboration graph; an indicator of whether the researcher is expanding their epistemic community or primarily operating within an existing one.

The c-index for author $v_i$ is defined as the average of their publication-specific scores:

$$C\ (v_i) = \frac{1}{|P(v_i)|} \cdot \sum_{p \in P(v_i)} C_{v_i,p}$$

Where:

- $P(v_i)$ is the set of publications authored by $v_i$.
- $C_{v_i,p}$ is the community score of $v_i$ for publication $p$.

Publication-Specific Score
For each publication $p$, the community score $C_{v_i,p}$ is the sum of the factors for each feature $f$:

$$C_{v_i,p} = \sum_{f \in F} C_{v_i,p,f}$$

Where:
- $F = \{gender, country, discipline\}$ is the set of features.
- $C_{v_i,p,f}$ is the community factor for feature $f$ in publication $p$.

Community Factor Calculation Per Feature
For each feature $f$ and publication $p$, the community factor $D_{v_i,p,f}$ is calculated differently depending on whether $f$ is binary (e.g., gender) or categorical (e.g., country, discipline).

Community Factor for Binary Features (e.g., gender)
Let:

- $C_p$ be the set of co-authors on publication $p$, excluding $v_i$;
- $N_p = |C_p|$ be the total number of co-authors on publication $p$.

We then define the similarity cost for a binary feature $f$ as:

$$Cost_{v_i,p,f}^{similar} = 1 + \sum_{v_j \in C_p} \delta f(f(v_i), f(v_j)) \cdot 1/\ bonus(v_j),$$

Where:

- $\delta f(f(v_i), f(v_j))$ is the Kronecker delta function (1 if features match, 0 otherwise),
- $bonus(v_j)$ is the novelty multiplier applied to collaborator $v_j$ (see below).=

-

The community factor for the binary feature $f$ for author $v_i$ on publication $p$ is then defined as:

$$D_{v_i,p,f} = \left(\frac{1}{Cost_{v_i,p,f}^{similar}}\right) \cdot N_p \cdot w_f$$

Where $w_f$ is the base weight for feature $f$.

Community Factor for Categorical Features (e.g., country, discipline)

Let:

- $V_f$ be the set of all possible categories for feature $f$,
- $f(v_j) \in V_f$ be the category of co-author $v_j$.

Then:

$$N_f^{(v)} = \sum_{v_j \in C_p, f(v_j)=v} bonus(v_j) \; for\; each\; v \in V_f$$

$$N_f^{Collaborators} = \sum_{v \in V_f} N_f \; (v) \text{ - w_f}$$

-
-

The categorical feature contribution is:

$$C_{v_i,p,f} = \left(\frac{1}{N_f \; (f(vi))} \cdot N_f^{collaborators}\right) \cdot |V_f| \cdot w_f$$

where:

- $f_{v_i}$ is the category of author $v_i$ for feature $f$.
- $|v_f|$ is the number of unique categories in $v_f$, representing the breadth of the community.

Capturing Collaborative Novelty

To model the temporal dynamics of scholarly collaboration, we introduce a *novelty multiplier* that adjusts the contribution of each co-author based on whether they represent a new connection in the author's co-authorship network.

For a given author $v_j \in C_p$, on publication *p,* we define the novelty multiplier as:

$$bonus(v_j) = \begin{cases} 1 + \delta, & if \deg_G(v_j) = 1 \\ 1, & otherwise \end{cases}$$

where:

- $deg_G(v_j)$ is the degree of node $v_j$ in the co-authorship graph $G$ constructed from the full publication history of $v_i$ representing the number of distinct collaborations between $v_i$ and $v_j$,
- $\delta \in \mathbb{R}^+$ is a tunable parameter controlling the strength of the novelty adjustment (e.g., $\delta = 0.8$).

This multiplier is applied to the co-author's contribution in the calculation of gender, geographic, and disciplinary diversity scores. By weighting new connections more heavily, the novelty multiplier emphasizes the expansion of an author's collaborative network and promotes engagement with a broader range of perspectives and research communities.

Data:

To demonstrate the potential applications of the c-index we created a sample dataset of authors, mimicking the output of the Dimensions database. The columns included were:

- publication ID,
- first name,
- last name,
- affiliated country,
- affiliated country code,
- and affiliated city.

For real-world usage, we could use gender information when available and predict the likely gender for unknown cases using predictive models, acknowledging the limitation that these models are predictive and operate on a male-female gender binary. The binary nature of these tools is a limitation because it does not capture the full representation of human embodiment and how people identify, and furthermore, risks mis-gendering authors based upon statistical, gendered assumptions about their name.[10] One way of addressing this limitation is to create best practices such as using Open Researcher and Contributor ID (ORCID) to uniquely identify an author and give them an opportunity to share the details of their identity.

**Results:**

In this section, we will walk through the formal definition of the c-index with examples. The schema used for the data is driven from the Dimensions Database[1] supplemented by a couple of new fields to better illustrate these examples.
*Please note that these attributes are used only to demonstrate the idea of c-index. Actual attributes may be different in real implementation.*

Binary Features:

We will start with the binary feature, with the next few examples focusing on *Gender*. The table below shows calculations for a single paper with 4 authors. Each row represents a reference author with their corresponding gender factor, based on other collaborators.

**Table 1: Gender Factor Calculation for Single Paper**

| pub_id | first_name | last_name | gender | # of "same" gender | # of collaborators | gender factor |
|---|---|---|---|---|---|---|
| 123 | Omar | Hassan | M | 3 | 3 | 1 |
| 123 | Daniel | Young | M | 3 | 3 | 1 |
| 123 | Joshua | Carter | M | 3 | 3 | 1 |
| 123 | Anna | Nguyen | F | 1 | 3 | 3 |

- Reference Author: It is important to note the c-index is calculated on an author-by-author basis, meaning the gender factor needs to apply to a specific author.

The calculation incorporates the proportion of the number of authors who share the same gender as the reference author to the total number of collaborators. An author with a higher proportion is considered less diverse. To simplify, we take the reciprocal of that proportion to measure relative community membership, such that a higher number reflects plurality.

In this case, the calculated relative score is called the gender factor.

To further illustrate the idea, the table below shows different calculated gender factors based on the various possible compositions of 4 different authors.

**Table 2: All Possible Gender Factors for a Paper with 4 Authors**

| Total authors | Number of male author(s) | Number of female author(s) | Individual male author gender factor | Individual female author gender factor |
|---|---|---|---|---|
| 4 | 4 | 0 | 0.75 | N/A |
| 4 | 3 | 1 | 1 | 3 |
| 4 | 2 | 2 | 1.5 | 1.5 |
| 4 | 1 | 3 | 3 | 1 |
| 4 | 0 | 4 | N/A | 0.75 |

Looking at the table above, a paper with 3 male authors and 1 female author gives a gender factor of 1 to each male author and a gender factor of 3 to the female author. We view the factors to be “equalizing” or “balancing.” In this context, the female author needs to be valued 3 times as much for equal representation. WiA research trajectory is always constrained by outside structural factors, and authors are of course not completely “free” to choose the composition of the teams they work with. However, by including this in the c-score index,the hope is that promoting plurality could lead to a larger, cross disciplinary community, and potentially increase the likelihood of its underrepresented members benefiting from it.

Categorical Features:

The following examples highlight a categorical feature, *Country* which represents the author's affiliated country. It extends the idea of binary features by taking into consideration the size of the category set.

**Table 3: Country Factor Calculation for a Single Paper**

| pub_id | first_name | last_name | country | number of countries | **country relative plurality** | **author relative plurality (country)** |
|---|---|---|---|---|---|---|
| 123 | Omar | Hassan | Egypt | 3 | 3 | 9 |
| 123 | Daniel | Young | United States | 3 | 1.5 | 4.5 |
| 123 | Joshua | Carter | Italy | 3 | 3 | 9 |
| 123 | Anna | Nguyen | United States | 3 | 1.5 | 4.5 |

In this case, the Country relative plurality is calculated the same way as the Gender factor. In addition, the total count of the category elements (in this case 3 for 3 distinct countries) is used to give weight to the number of different countries represented.
This is required because the relative plurality calculation doesn’t take into account the total number of different countries represented. As shown in the below example, we see that the author from Egypt has the same ‘country relative plurality’, even when all the remaining authors are from the same country (United States).

**Table 4: Country Factor Calculation for a Single Paper (Modified)**

| pub_id | first_name | last_name | country | number of countries | **country relative plurality** | **author relative plurality (country)** |
|---|---|---|---|---|---|---|
| 123 | Omar | Hassan | Egypt | 2 | 3 | 6 |
| 123 | Daniel | Young | United States | 2 | 1 | 2 |
| 123 | Joshua | Carter | United States | 2 | 1 | 2 |
| 123 | Anna | Nguyen | United States | 2 | 1 | 2 |

- The country in the 3rd row has been changed from Italy to the United States and we expect the author's relative plurality to change. But we see that the author from Egypt has the same ‘country relative plurality’ as in Table 3.

While incorporating the size of the category set makes the calculation more precise, it also tends to inflate the overall factor, if there is a large number of elements in the set. To counter this, we propose a notion of *Base Factor* that can be defined per feature. This will help normalize the value of different features.

For categorical features, there is another nuance, where every element within that category may not be weighed the same. For argument’s sake, let’s say the US should weigh half as much as other countries. The below table shows this idea by extending Table 3, with an additional *Country Weight* column.

**Table 5: Country Factor Calculation for a Single Paper (With Weight)**

| pub_id | first_name | last_name | country | number of countries | country weight | **country relative plurality** | **author relative plurality (country)** |
|---|---|---|---|---|---|---|---|
| 123 | Omar | Hassan | Egypt | 3 | 1 | 3 | 9 |
| 123 | Daniel | Young | United States | 3 | 1/2 | 1.5 | 2.25 |
| 123 | Joshua | Carter | Italy | 3 | 1 | 3 | 9 |
| 123 | Anna | Nguyen | United States | 3 | 1/2 | 1.5 | 2.25 |

The c-index Calculation:

Now that we have gone over the calculation for *Relative Plurality Factor* for different types of features, we now focus on the c-index calculation, for a reference author. To get the c-index, we first calculate the *Paper Factor*, which is derived by simply summing up the individual feature factors for a given paper and author combination.

See below example in Table 6; it uses 3 different features

- Country
- Gender
- Field (specialization)

Then the various Paper Factors for a given reference author is averaged across all their papers. This value is the c-index for that author.

**Table 6: Paper Factor & c-index Calculation**

| Pub id | Name | Country | Gender | Field | Country factor | Gender factor | Field factor | **Paper factor** | **c-index** |
|---|---|---|---|---|---|---|---|---|---|
| 246 | Adam | Italy | M | Health | 20 | 1.66 | 10 | 31.66 | 27.55 |
| 789 | Adam | Italy | M | Health | 12 | 1.5 | 12 | 25.5 | |
| 369 | Adam | Italy | M | Health | 12 | 1.5 | 12 | 25.5 | |
| 246 | David | US | M | Health | 6.66 | 1.66 | 10 | 18.32 | 18.32 |
| 246 | Emily | Cuba | F | CS | 20 | 1.66 | 10 | 31.66 | 27.55 |
| 789 | Emily | Cuba | F | CS | 12 | 1.5 | 12 | 25.5 | |
| 369 | Emily | Cuba | F | CS | 12 | 1.5 | 12 | 25.5 | |
| 246 | Maria | Mexico | F | SS | 20 | 1.66 | 20 | 41.66 | 30.89 |
| 789 | Maria | Mexico | F | SS | 12 | 1.5 | 12 | 25.5 | |
| 369 | Maria | Mexico | F | SS | 12 | 1.5 | 12 | 25.5 | |
| 246 | Robert | US | M | Biology | 6.66 | 1.66 | 20 | 28.32 | |
| 789 | Robert | US | M | Biology | 12 | 1.5 | 12 | 25.5 | 26.44 |
| 369 | Robert | US | M | Biology | 12 | 1.5 | 12 | 25.5 | |
| 246 | Sophia | US | F | CS | 6.66 | 1.66 | 10 | 18.32 | 18.32 |

While the c-index rewards plurality across gender, geography, and discipline, it also recognizes that the structure of collaboration itself plays a critical role in shaping diversity. In academic practice, long-term partnerships are often essential to sustaining productivity, mentorship, and trust. Yet such stability, however valuable, may inadvertently limit the entry of new collaborators and perspectives over time. To account for this, the model introduced a *novelty multiplier*, an adjustment that increases the contribution of co-authors who appear for the first time in a given author's collaboration network. This mechanism emphasizes the value of intellectual openness and epistemic renewal, rewarding scholars who continue to expand the boundaries of their research communities. By foregrounding the formation of new ties, the c-index favors not only diverse teams but also inclusive practices that bring new voices into ongoing scholarly conversations.

Table 7 demonstrates how new co-authorships contribute to the final c-index score through a *novelty multiplier*, which emphasizes the importance of forming new collaborative ties. For each publication, a bonus factor is applied to co-authors who appear for the first time in the focal author's collaboration network. The multiplier amplifies the diversity contribution of these novel collaborators across all three dimensions: country, gender, and discipline. Authors who repeatedly co-publish with the same group continue to accumulate diversity contributions, but without the added multiplier effect. In doing so, the metric encourages both sustained pluralism and the expansion of scholarly networks.

**Table 7: Paper Factor & c-index Calculation with novelty multiplier**

*New Bonus: 80%*

| | | | | | | **Calculations after applying the bonus function** | | | | |
|---|---|---|---|---|---|---|---|---|---|---|
| **Pub id** | **Name** | **Country** | **Gender** | **Field** | **c-index** | **Country factor** | **Gender factor** | **Field factor** | **Paper factor** | **c-index** |
| 246 | Adam | Italy | M | Health | 27.56 | 12 | 1.5 | 12 | 25.5 | 29.6 |
| 789 | Adam | Italy | M | Health | 27.56 | 26.4 | 1.96 | 9.43 | 37.79 | 29.6 |
| 369 | Adam | Italy | M | Health | 27.56 | 12 | 1.5 | 12 | 25.5 | 29.6 |
| 789 | David | United States | M | Health | 18.33 | 7.83 | 2.37 | 12.86 | 23.05 | 23.05 |
| 246 | Emily | Cuba | F | Computer Sci | 27.56 | 12 | 1.5 | 12 | 25.5 | 29.6 |
| 789 | Emily | Cuba | F | Computer Sci | 27.56 | 26.4 | 1.96 | 9.43 | 37.79 | 29.6 |
| 369 | Emily | Cuba | F | Computer Sci | 27.56 | 12 | 1.5 | 12 | 25.5 | 29.6 |
| 246 | Maria | Mexico | F | Social Sci | 27.5 | 12 | 1.5 | 12 | 25.5 | 33.49 |
| 789 | Maria | Mexico | F | Social Sci | 27.5 | 26.4 | 1.96 | 26.4 | 54.76 | 33.49 |
| 369 | Maria | Mexico | F | Social Sci | 27.5 | 12 | 1.5 | 12 | 25.5 | 33.49 |
| 111 | Maria | Mexico | F | Social Sci | 27.5 | 21.6 | 1.89 | 4.7 | 28.19 | 33.49 |
| 246 | Robert | United States | M | Engineering | 26.44 | 12 | 1.5 | 12 | 25.5 | 28.37 |
| 789 | Robert | United States | M | Engineering | 26.44 | 5.74 | 1.96 | 26.4 | 34.1 | 28.37 |
| 369 | Robert | United States | M | Engineering | 26.44 | 12 | 1.5 | 12 | 25.5 | 28.37 |
| 789 | Sophia | United States | F | Computer Sci | 18.33 | 7.83 | 2.37 | 12.86 | 23.05 | 23.05 |

## Discussion

Although plurality is a multidimensional, qualitative construct that is impossible to fully quantify, we strove to include some of the most salient features of plurality, namely both the demographic characteristics and epistemic plurality of authors.[11] By doing so, the c-index is able to reveal important patterns of collaboration across different genders (and also enable further research on how gender inequalities intersect with racial inequalities), disciplines, and geographical regions. Traditional metrics like the h-index often overlook these dimensions, focusing instead on individual output and citation counts without accounting for the plurality of contributions or collaborations. The c-index seeks to address these gaps by providing a multidimensional view of scholarly impact, highlighting both gender, country, and disciplinary plurality as essential components of academic evaluation.

Diversity continues to be a significant concern in academic research. Abdalla et al. (2023) demonstrated that both JAMA and NEJM featured between 0.3% and 0.9% American Indian/Alaska Native (AI/AN) first authors since 1990, with a slope of zero over the last 30 years.[12] This sobering percentage is far less than the actual proportion of AI/AN people in the US, which is 1.7%. Similarly, women of color face the lowest authorship rates, as evidenced by the mere 0.3% to 3.6% representation of Black women as lead or senior authors in various journals. Likewise, Latina researchers encounter limited opportunities, with their lead or senior authorship ranging from 0.4% to 2.5%.[13] These statistics underscore systemic barriers faced by underrepresented groups in academic publishing. JAMA and NEJM were chosen as examples because they are among the most prestigious medical journals, and their trends reflect broader patterns in academic publishing. The lack of progress in plurality at these journals illustrates a larger issue of inequitable representation across the field. These disparities highlight the urgency of developing tools, such as the c-index, that place explicit value on plurality in research collaborations, both to improve inclusivity and to ensure that academic publishing reflects the representativeness of the populations it serves.

The proposed c-index is designed to evaluate scholarly impact while offering a more nuanced picture of academic work. It incorporates multiple axes of plurality, including gender, country and disciplinary specialty, which can all be weighted according to the availability of data. For example, the c-index can account for home institution ranking as a proxy for an author's access to resources and visibility, while interdisciplinarity captures the extent of cross-specialty collaboration within a paper. Specialty representation highlights the inclusion or exclusion of underrepresented specialties, drawing attention to areas of research that may lack diverse perspectives. To make the system transparent, the weighting parameters for various factors—such as first and senior authorship positions—are adjustable, allowing the metric to evolve alongside the data available and the plurality goals of academic institutions. While the c-index is not prescriptive, it provides a flexible framework for incorporating these dimensions into the evaluation of scholarly impact.

Despite its ambitions, the c-index cannot and does not claim to encapsulate the full complexity of diversity, collaboration, or epistemic pluralism. As with any metric, it reflects a deliberate act of abstraction, one that privileges what is measurable over what is socially embedded or contextually nuanced. This is not a failure of the metric, but rather a fundamental epistemological constraint shared by all attempts to quantify complex social constructs such as pluralism or scholarly community. As scholars in science and technology studies have long argued, indicators are never neutral instruments[17-20]. Rather, they are the result of socio-technical choices; inherently normative tools that encode assumptions about what counts as valuable knowledge and who is authorized to produce it. The c-index, in this sense, should be understood not as a mirror of collaboration but as a constructed lens shaped by an explicit commitment to inclusivity across demographic, geographic, and disciplinary lines. Its limitations are not only technical, but ontological, reminding us that any effort to quantify the social fabric of science risks omitting the very power structures, exclusions, and histories it seeks to challenge. Rather than offering a thorough perspective of scholarly excellence, the c-index proposes a new axis of visibility, one that brings attention to the composition of teams, not just their citation counts[16,17]. It is a value-laden intervention meant to provoke institutional reflection, not a final measure of academic merit.

The c-index offers several advantages over traditional metrics. It complements the h-index by not only evaluating an individual's scholarly output but also encouraging diverse, interdisciplinary, and collaborative research. Research in psychology has long shown a tension between extrinsic rewards (e.g., publication counts, citation-based metrics) and intrinsic motivations (e.g., curiosity, purpose, and personal growth) that sustain meaningful academic work[42,43]. The kind of plural and inclusive knowledge production the c-index aims to foster is not only beneficial for scientific innovation but also conducive to more reflective, socially attuned, and intellectually generative academic lives. As a scholarly community, academic institutions must work toward both recognizing these intrinsic rewards and avoiding systems of evaluation that crowd them out. When left unbalanced, metrics like the h-index risk incentivizing strategic, self-interested behavior that may erode the interpersonal and epistemic value of collaboration. The c-index offers a complementary orientation, one that aligns institutional reward structures with the deeper internal commitments many scholars already find most meaningful. In practical terms, funding agencies could use the c-index to prioritize projects led by diverse research teams, while universities could consider it as part of faculty evaluations for promotion and tenure, ensuring that contributions to diversity, equity, and inclusion (DEI) efforts are recognized alongside traditional measures of academic success.[14] Importantly, the c-index does not diminish the value of scientific excellence but instead adds community membership as an additional layer to academic assessment. By doing so, it enhances the recognition of diverse perspectives without compromising the rigor or quality of research.

However, there are challenges and limitations to consider. One potential risk is the formation of citation teams, where researchers repeatedly collaborate with the same diverse team to artificially inflate their c-index.[15] While such teams may appear inclusive, this pattern can dilute the intent of the c-index, which aims to encourage genuine plurality through expanding collaborative networks. To address this, the c-index incorporates a novelty multiplier that progressively reduces the contribution of repeated co-authors across feature dimensions. This adjustment rewards authors who engage with new collaborators over time, reinforcing the value of expanding diversity not just in demographics but in network structure as well, thereby promoting broader inclusivity in research.

Another limitation is the potential misuse of the system to include "placeholder" authors—individuals added to author lists for the sake of appearing diverse without contributing meaningfully to the work.[16] Although the c-index cannot fully prevent such practices, a robust peer review process and transparency in authorship contributions can help address this issue. Additionally, critics will argue that metrics like the c-index are more complex than traditional metrics, but this complexity reflects the inherent challenges of quantifying plurality. By making the factors and calculations public, the c-index encourages scholars to become active participants in DEI efforts, fostering a culture of awareness and accountability.

While the c-index advances a multidimensional approach to plurality, it currently omits one of the most consequential axes of inequality in academia: race. This exclusion is not due to lack of relevance but rather reflects the methodological and ethical complexities associated with incorporating racial and ethnic identity into algorithmic metrics. Unlike features such as discipline or institutional affiliation, which are relatively stable and widely available in bibliometric data, racial identity is rarely recorded in publication metadata, and attempts to infer it through proxies such as surnames, geolocation, or photographs risk reinforcing forms of epistemic and representational harm. Moreover, self-identification, the gold standard in race data collection, is often absent from global research infrastructures and is complicated by cultural, political, and legal variations in how race is defined, perceived, and operationalized across countries[31,44]. Still, to ignore race entirely is to overlook one of the most deeply entrenched dimensions of scholarly exclusion, including in patterns of authorship, citation, and gatekeeping[32-36]. That said, race is not the only demographic or identity-based factor that poses these challenges. Other dimensions, such as gender identity (beyond the binary), Indigeneity, language, or disability, are similarly underrepresented or misrepresented in academic metadata, and their salience often varies depending on the research field, topic, or geography[37-39]. These exclusions are not technical oversights, but structural limitations of the systems that define what and who is visible in academic production[26]. As the literature on race and AI has shown, including identity categories in technical systems is both indispensable and deeply contested, precisely because structures of inequality shape who gets to generate, circulate, and benefit from knowledge[40,41]. We raise this as a critical area for future development: any effort to embed race or other identity markers into the c-index must proceed with care, centering the perspectives of scholars from historically excluded communities, and avoiding reductive or essentialist codifications.

The current iteration of the c-index, then, is an incomplete instrument, one that highlights the necessity of building fairer academic infrastructures, while also revealing the epistemological limits of measurement. Yet its value lies not in offering a universal solution to structural inequality, but in its ability to adapt to different contexts of plurality, whether demographic, disciplinary, or epistemic. The c-index does not prescribe a one-size-fits-all approach to plurality. For instance, even a theoretically all-white author group collaborating in Sweden on a healthcare delivery research article could achieve a high c-index by incorporating stakeholders from a wide array of fields, such as health economics, policy, and clinical care. This flexibility ensures that the metric accommodates the unique needs of different research contexts while emphasizing the value of interdisciplinary and inclusive scholarship. Ultimately, the c-index aims to ensure that researchers draw from a range of specializations, identities, social positions and perspectives to offer unique insights into medical issues and their broader social contexts.[17] By addressing the limitations of traditional metrics and incentivizing plurality, the c-index represents a meaningful step toward a more equitable research landscape.

While the c-index introduces a new axis of scholarly assessment grounded in plurality, it must be recognized as only one component of a much broader, systemic transformation that academic research urgently needs. Metrics alone cannot repair structural inequalities embedded in the global knowledge economy; from exploitative labor practices, such as the rise of precarious adjunct contracts and the unpaid labor of early-career scholars[21], to exclusionary barriers like exorbitant article processing charges that perpetuated existing disparities[22-24]. The very infrastructures of data access, authorship, and funding remain overwhelmingly controlled by institutions in the Global North, reinforcing unequal partnerships under the guise of collaboration[25, 26]. These inequities are further entrenched by academic gatekeepers; senior, well-resourced, majority-male scholars, who disproportionately occupy editorial boards, grant panels, and markers of institutional prestige[27-30]. As such, any attempt to realign evaluation criteria, including through plurality-focused indices, must be paired with meaningful structural reforms in funding models, data governance, institutional labor practices, and international research ethics. What we offer in this paper is not a comprehensive fix, but a recalibration of one small but influential mechanism - the metric - which, as seen with the h-index, has had outsized impact on career trajectories, often at the expense of equity. Recognizing the c-index as part of a multipronged strategy, rather than a standalone solution, is essential to situating this work within a broader agenda for reform.